\documentclass[
	nofootinbib,
        amsmath,
        amssymb,
        superscriptaddress,
         aps,
         prb,
         longbibliography,
         floatfix,
         twocolumn
]{revtex4-2}
\usepackage[english]{babel}
\usepackage[utf8]{inputenc} 
\usepackage{dsfont}
\usepackage{SIunits}
\usepackage[caption=false,label font=bf,labelformat=simple]{subfig}
\usepackage{graphicx}     % Allows to insert images from outside
\usepackage{color}
\usepackage[normalem]{ulem}
\usepackage{soul}
\usepackage{cancel}
\usepackage{hyperref}% add hypertext capabilities
\usepackage{slashed}
\usepackage[normalem]{ulem}
\usepackage{verbatim}

\begin{document}
\title{Charging energy effects on a single-edge anyon braiding detector}
\author{Noé Demazure}
\affiliation{Aix Marseille Univ, Universit\'e de Toulon, CNRS, CPT, Marseille, France}
\author{Flavio Ronetti}
\affiliation{Aix Marseille Univ, Universit\'e de Toulon, CNRS, CPT, Marseille, France}
\author{Benoît Grémaud}
\affiliation{Aix Marseille Univ, Universit\'e de Toulon, CNRS, CPT, Marseille, France}
\author{Laurent Raymond}
\affiliation{Aix Marseille Univ, Universit\'e de Toulon, CNRS, CPT, Marseille, France}
\author{Masayuki Hashisaka}
\affiliation{Institute for Solid State Physics, University of Tokyo, 5-1-5 Kashiwanoha, Kashiwa, Japan}
\author{Takeo Kato}
\affiliation{Institute for Solid State Physics, University of Tokyo, 5-1-5 Kashiwanoha, Kashiwa, Japan}
\author{Thierry Martin}
\affiliation{Aix Marseille Univ, Universit\'e de Toulon, CNRS, CPT, Marseille, France}

\begin{abstract}
We investigate the influence of capacitive coupling on the detection of anyon braiding in a single-edge interferometer realized in the fractional quantum Hall regime. In this setup, a quantum point contact bends a single edge into a loop, where tunneling occurs at the open end and is controlled by the QPC voltage. In contrast with previously studied two-edge geometries, the weak backscattering regime is dominated by the first-order perturbative term, allowing quantum transport quantities to factorize into a non-universal prefactor and a braiding-induced contribution that provides direct access to the universal statistical angle $\pi\lambda$. While previous analyses neglected edge-to-edge capacitance, we show that capacitive effects, which are known to play a crucial role in mesoscopic capacitors, modify both the current and the current cross-correlations. Using a two-point Green’s function formalism augmented by Dyson’s equation to include the charging energy, we quantify how the fluctuations of the cross-correlations depend simultaneously on $\lambda$ and on the capacitance of the loop. Our results indicate that a reliable extraction of the statistical angle requires a parallel measurement of the loop capacitance, which can be implemented via a charged gate coupled to the junction.  
\end{abstract}

\newcommand{\fr}[1]{\textcolor{red}{#1}}
\newcommand{\nd}[1]{\textcolor{blue}{#1}}
\newcommand{\lr}[1]{\textcolor{green}{#1}}
\newcommand{\sgn}[0]{\ensuremath{
\,\mbox{sgn}
}}

\maketitle
\section{Introduction}
Due to their conceptual richness and potential applications in quantum information, anyonic statistics have become a central theme in condensed matter physics, particularly in the study of the fractional quantum Hall effect \cite{Tsui82,Laughlin83}. 
In his seminal work, Laughlin explained the appearance of fractional conductance plateaus by proposing that the low-energy excitations of the quantum Hall fluid are quasiparticles carrying a fraction of the electron charge. 
These excitations were soon recognized to obey fractional exchange statistics \cite{Leinaas77,Wilczek82,Arovas84}, such that the many-body wave function acquires a phase factor \( e^{i\pi\lambda} \), with \( 1/\lambda \in 2\mathbf{N}+1 \), upon exchanging two quasiparticles. 

Quasiparticle charge has been probed through transport measurements using quantum point contacts (QPCs), where two edges of a Hall droplet are brought sufficiently close by metallic gates to allow tunneling. 
The analysis of current fluctuations in a voltage-biased QPC provides access to the effective charge \( e^* \) via the Fano factor, leading to direct measurements \cite{Kane94,Saminadayar97,dePicciotto97,Crepieux04,Martin05,Biswas22}, later corroborated by complementary methods \cite{Goldman95,JensMartin04,Kapfer19,Bisognin19}. 
In contrast, detecting fractional statistics has proven substantially more challenging: although numerous interferometric and tunneling-based proposals have been formulated over the years \cite{Chamon97,Safi01,Vishveshwara03,Law06,Hou06,Bishara08,Halperin11,Campagnano12,Rosenow12,Levkivskyi12,Campagnano13,Lee19,Carrega21}, only very recently have the first convincing experimental signatures been reported.

The first convincing measurements of fractional statistics were achieved using the Fabry–Pérot interferometer~\cite{Nakamura20,Chamon97,Halperin11,McClure12,Carrega21,Camino05,Roosli20,Nakamura19,Ronen21,Nakamura23,Kim24,Kim24b,Werkmeister24,Samuelson24,Werkmeister25} and the anyon collider~\cite{Bartolomei20,Rosenow16,Han16,Lee22,Morel22,Mora22,Schiller23,Jonckheere23}. In the Fabry–Pérot setup, two QPCs connected by edge channels form a coherent loop in which bulk anyons entering or leaving produce discrete phase slips of $\pi\lambda$, though the timing of these events is not precisely controllable. In the anyon collider, two injection QPCs generate trains of anyons that collide at a central QPC, producing braiding-induced contributions in the output currents and cross-correlations. These allow extraction of the statistical angle $\pi\lambda$, provided the scaling dimension $\delta$ is known, although its experimental determination can be affected by environmental renormalization~\cite{Snizhko15,Rech20,Ebisu22,Schiller22,Bertin23,Iyer23,Acciai24,Veillon24,Ramon25,Schiller24,Ruelle24,Braggio12}. Another approach implements a Mach-Zehnder–type interferometer based on co-propagating interface modes~\cite{Batra2023,Ghosh2024,Batra2025}, where localized quasiparticles can be induced by a top gate, producing flux-dependent phase shifts and anyonic exchange effects across various fractional fillings.

Recently, a novel type of interferometer has been proposed \cite{Ronetti25,Ronetti25b}. 
In this setup, a loop is formed along a single edge, which is bent into a loop by a QPC.  Tunneling occurs at the open end of this loop and it is controlled by the voltage gate of the QPC. 
This constitutes a highly asymmetric Fabry-Perot interferometer where propagation is chiral (and ideal) in one arm, along the perimeter of the loop, and where (weak) anyon tunneling can occur in both directions in the other arm.

This is also precisely the same circuit employed to study the universality of the relaxation resistance of the mesoscopic capacitor \cite{buttiker93,pretre96}, which was later demonstrated experimentally in the integer quantum Hall effect regime \cite{gabelli06}. % TM
This single-edge configuration is in contrast with previously studied two-edges configurations. Indeed, in the weak backscattering regime, the first-order perturbative term is the leading one; in this case, quantum transport quantities exhibit a clear factorization between a non-universal pre-factor and a braiding-induced one which grants direct access to the universal statistical angle $\pi\lambda$.  

This analysis was initially performed within a minimal model neglecting the edge-to-edge capacitance. Yet this effect is known to play a crucial role in mesoscopic capacitors \cite{Gabelli07,Parmentier12}, whose geometries closely resemble that of the present interferometer. The goal of this work is therefore to evaluate the robustness of the extraction of the statistical parameter $\lambda$ in the presence of such capacitive coupling. Fortunately the generalization of the mesoscopic capacitor which includes exactly (within the Luttinger model) charging effects is readily available
\cite{hamamoto10,mora10}.
To this end, we revisit the calculation using the same two-point Green’s function formalism, now modified through Dyson’s equation to include a charging energy. This modification alters the expressions for both the current and the current cross-correlations. The fluctuations of the latter remain sensitive to the value of $\lambda$, but also depend on the capacitance of the loop. Consequently, a simultaneous measurement of the capacitance should be incorporated into the experimental protocol, which we propose to perform by coupling a charged gate to the junction.

The remainder of the article is organized as follows. 
In Section~\ref{sec:Model}, we introduce the main elements of the model, including bosonization, the Hamiltonian of the $\Omega$-junction, and the input states. 
Section~\ref{sec:GF} incorporates the effect of the charging energy into the Green's functions of the fractional quantum Hall edge via Dyson's equation. 
These Green's functions are then employed in Section~\ref{sec:Current} to compute the edge current to zeroth and first order in the tunneling amplitude. 
Cross-correlations are analyzed in Section~\ref{sec:Noise} using a refined model of the injection QPC. 
Section~\ref{sec:Gate} extends the setup by including a gate, whose effect is fully incorporated in the calculation. Finally in Section~\ref{sec:Conclusions} we draw our conclusions. Two appendices contain the details of our calculations. Throughout, we adopt units in which $\hbar=k_B=1$ and the electron charge is $-e<0$.

\section{Model \label{sec:Model}}

\begin{figure}
    \centering
    \includegraphics[scale=.2]{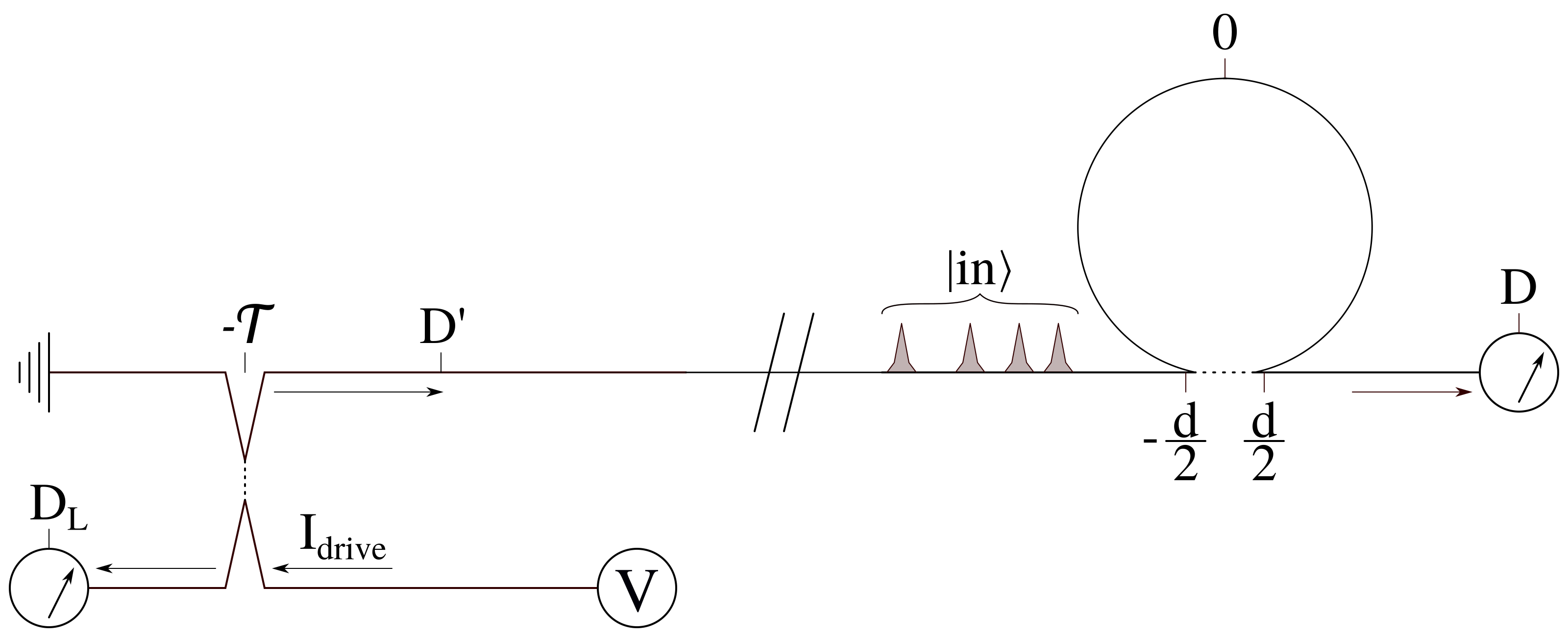}
    \caption{Schematic representation of the setup showing positions on the $x$ axis. On the left, the injection QPC is driven by a constant voltage and emits a tunneling current on the right-movers edge. Far after, that current is represented as a sequence of quasi-particles in a certain state $|in\rangle$. They travel the loop, influencing the tunneling process between $-d/2$ and $d/2$. Please note that the $x$ axis is curvilinear, $d$ corresponding to the length of the loop following the upper path. The distance separating the points $-d/2$ and $d/2$ by the QPC is approximately 0. The current is measured by a detector situated at a position $D>d/2$.}
    \label{schOm}
\end{figure}
We consider a FQHE bar at the Laughlin filling factor $\nu = 1/(2n+1)$, with $n \in \mathbb{Z}$. In this regime, the bulk is insulating, and its excitations are quasiparticles carrying a fractional charge $-e^* = -e\nu$ and obeying fractional statistics characterized by an exchange angle $\pi \lambda = \pi \nu$. Conduction occurs only along the edges of the bar, where a single chiral channel emerges. In the following, we focus on a single edge. Since our model involves only one edge mode, we set its velocity $v = 1$, as chirality allows us to treat space and time on an equal footing. In the Tomonaga-Luttinger theory \cite{Kane94,Wen95,vonDelft98}, a boson field $\phi$ is used to describe the single edge mode, whose dynamics can be captured by the Wen Hamiltonian \cite{Wen95}
\begin{equation}
    H_W=\frac{1}{4\pi}\int dx\;\left[\partial_x\phi(x)\right]^2.
\end{equation}
The density of charge can also be expressed in terms of the boson field $\phi$ as
\begin{equation}
\label{eq:density}
    \rho(x)=-e\frac{\sqrt{\nu}}{2\pi}\partial_x\phi(x).
\end{equation}
To probe the statistical angle of anyons, our protocol relies on a controlled deformation of the edge that forms a localized droplet of total perimeter $d$. As shown in Fig.~\ref{schOm}, this deformation creates an $\Omega$-shaped junction where quasi-particles can tunnel between two distinct edge segments. We introduce a curvilinear coordinate $x$ along the edge, with $x = 0$ at the center of the loop. The tunneling processes occur within the junction region, between $x = -d/2$ and $x = d/2$. We refer to this region as the \emph{junction}, and to the portion 
$-d/2 < x < d/2$ as the \emph{loop}. In the limit where the overlap of the edge wave functions at $x = -d/2$ and $x = d/2$ is weak, the region between these two points is occupied by the FQHE bulk, allowing only fractional quasiparticles to tunnel through. In this regime, tunneling across the junction can be modeled by
\begin{equation}
    H_T=\Gamma\sum_{\epsilon=\pm} e^{-i\epsilon\kappa} \psi^{\dagger}\left(\epsilon\frac{d}{2}\right)\psi\left(-\epsilon\frac{d}{2}\right)
\end{equation}
where $\Gamma$ is the tunneling amplitude in modulus, $\kappa$ the phase acquired by a quasiparticle after completing one turn around the loop and $\psi(x)$ is the quasi-particle annihilation operator at the position $x$. The phase $\kappa$ is due to the geometric difference of path and the Aharonov-Bohm phase induced by the external magnetic flux. The annihilation operator associated to quasi-particles can be expressed in terms of the boson field $\phi$ as
\begin{equation}
    \psi(x)=\frac{e^{i\sqrt{\nu}\phi(x)}}{\sqrt{2\pi a}}
\end{equation}
$a$ is the spatial cut-off of the theory. We neglect the presence of Klein factors in the quasi-particle operator, since they always drop out in the number-conserving correlation functions employed in our calculation~\cite{Guyon2002}. By using this expression, the bosonized version of the tunneling Hamiltonian becomes
\begin{equation}
    H_T=\frac{\Gamma}{2\pi a}\sum_\epsilon e^{-i\epsilon\kappa}e^{i\sqrt{\nu}\phi\left(\epsilon\frac{d}{2}\right)-i\sqrt{\nu}\phi\left(-\epsilon\frac{d}{2}\right)}.
\end{equation}
In the following, we will be interested in the effects of the interaction between charges inside the loop and how this can affect the detection scheme of the statistical angle introduced in Refs.~\cite{Ronetti25,Ronetti25b}. The latter can be included in our model by considered the following capacitive coupling
\begin{equation}
    H_E=\frac{1}{2C}\left[\int_{-\frac{d}{2}}^{\frac{d}{2}}dx\;\rho(x)\right]^2.
\end{equation}
The bosonized form of the above expression in terms of the density operator in Eq.~\eqref{eq:density} is
\begin{equation}
\label{eq:capacity}
    H_E=\frac{c_E}{4\pi d}\left[\phi\left(\frac{d}{2}\right)-\phi\left(-\frac{d}{2}\right)\right]^2,
\end{equation}
where we introduced the charging energy
\begin{equation}
\label{eq:charging_energy}
c_E\equiv\frac{e^2\nu d}{2\pi C}.
\end{equation}

Finally, we want to consider the time-domain braiding of a stream of incoming quasi-particles with the quasi-particles and quasi-holes created at the junction. At this purpose, we represent a sequence of $N$ quasi-particles arriving from the left with respect to the junction by using the input state formulated in terms of the time evolution operator $T$ :
\begin{equation}\begin{split}
    |in\rangle&=T\Bigl(e^{-i\sqrt{\nu}\sum_n\phi(-\mathcal{T},-\mathcal{T}+\tau_n)}\Bigr)|0\rangle
\end{split}\end{equation}
obtained from the ground state $|0\rangle$. The variables $-\mathcal{T}$ and $\tau_{n}$ are the coordinate of the injection QPC and the instants of reaching the middle of the loop, respectively. They are ordered such that $\tau_i<\tau_j\Leftrightarrow i<j$. Finite width is not considered in our work, its effects being insignificant for Laughlin fractions \cite{Ruelle23,Iyer24,Thamm24}.

Moreover, one can use the Keldysh ordering operator (see \cite{Martin05}) to express the evaluation of a product of operators
\begin{equation}\begin{split}
    &\langle in|T_K\Bigl[O_a(t_a^{\eta_a})O_b(t_b^{\eta_b})...\Bigr]|in\rangle=
    \\&\langle0|T_K\Bigl[e^{-i\sqrt{\nu}\sum_n\sum_{\zeta=\pm}\zeta\phi(-\mathcal{T},(-\mathcal{T}+\tau_n)^\zeta)}O_a(t_a^{\eta_a})O_b(t_b^{\eta_b})...\Bigr]|0\rangle
\end{split}\end{equation}
under the condition that $-\mathcal{T}$ is situated before any other position and time of the problem. This is automatically true in the limit where the injection QPC is situated either far before the loop. Else, it is still possible to enforce this condition using a multiple branch Keldysh contour \cite{kogan91} to generate the prepared anyon injection.

% stat here ?

\section{Green's functions \label{sec:GF}}
We aim to compute the transport properties of the tunneling junction in the presence of the capacitive coupling introduced in Eq.~\eqref{eq:capacity}. This is accomplished by deriving the Green's functions incorporating the capacitive coupling $H_E$ to all orders while handling the tunneling term $H_T$ perturbatively. These functions will serve as the fundamental building blocks for the transport calculations presented in the remainder of this paper.

Our analysis is most conveniently performed within the Keldysh formalism~\cite{Martin05}, where the bosonic Green's function is defined as
\begin{equation}
\begin{split}
  & G^{\eta_1\eta_2}(x_1,t_1;x_2,t_2) \equiv 
    \langle 0| T_K \bigl[\phi(x_1,t_1^{\eta_1}) \phi(x_2,t_2^{\eta_2})\bigr] |0\rangle \\
    &- \frac{1}{2}\langle 0| T_K \bigl[\phi(x_1,t_1^{\eta_1})^2\bigr] |0\rangle 
    - \frac{1}{2}\langle 0| T_K \bigl[\phi(x_2,t_2^{\eta_2})^2\bigr] |0\rangle,
\end{split}
\label{eq:defG}
\end{equation}
where $\eta_i = \pm$ specifies the branch of the Keldysh contour, ensuring proper time ordering in the out-of-equilibrium setting. The time evolution of the bosonic operators includes the effects of the Hamiltonian $H_E$, but not those of the tunneling Hamiltonian $H_T$.

A central object in the following analysis is the Green's function matrix in the $a/r/K$ basis, defined as
\begin{equation}
\check{G} \equiv
\begin{pmatrix}
0 & G^a \\
G^r & G^K
\end{pmatrix}
=
\begin{pmatrix}
0 & G^{+-} - G^{--} \\
G^{-+} - G^{--} & G^{-+} + G^{+-}
\end{pmatrix},
\end{equation}
with the inverse relation
\begin{equation}
G^{\eta_1\eta_2} = \frac{1}{2} \Bigl( G^K + \eta_1 G^a + \eta_2 G^r \Bigr).
\label{eq:MatrixKeldysh}
\end{equation}
We notice that the above expression entails the following useful identities
\begin{equation}
\label{eq:UsefulIdentity1}
\sum_{\eta_{1/2}}\eta_{1/2}G^{\eta_1\eta_2} = G^{a/r}
\end{equation}
and
\begin{equation}
\label{eq:UsefulIdentity2}
\sum_{\eta_1\eta_2}\eta_1\eta_2G^{\eta_1\eta_2} = 0.
\end{equation}
To obtain the full Green's functions in the presence of the capacitive coupling, we adopt a Dyson equation approach, starting from the bare Green's functions associated with the Hamiltonian $H_W$, denoted by $\check{G}_0$. The bare advanced Green's function is
\begin{equation}
G_0^a(x_1,t_1;x_2,t_2) = -i \pi \, \theta(t_2-t_1) \, \sgn(t_2-t_1 - x_2 + x_1),\label{eq:bareadvanced}
\end{equation}
while the bare retarded Green's function reads
\begin{equation}
G_0^r(x_1,t_1;x_2,t_2) = -i \pi \, \theta(t_1-t_2) \, \sgn(t_1-t_2 - x_1 + x_2).
\label{eq:bareretarded}
\end{equation}
The corresponding bare Keldysh component is
\begin{equation}
G_0^K(x_1,t_1;x_2,t_2) = 2\, \mathcal{G}_0(t_2 - t_1 - x_2 + x_1),
\end{equation}
with the auxiliary function
\begin{equation}
\mathcal{G}_0(t) = \ln \left| \frac{\sinh( i a \pi \Theta )}{\sinh( i a \pi \Theta - t \pi \Theta )} \right|,
\end{equation}
where $\Theta$ denotes the temperature.

To account for the effects of the charging energy, we introduce the self-energy $\check{\Sigma}$ derived from $H_E$:
\begin{equation}\begin{split}\label{Sigma}
    & \check{\Sigma}(x_1,t_1;x_1',t_1') = -i \frac{c_E}{2\pi d} 
    \Bigl[\delta(x_1-d/2)-\delta(x_1+d/2)\Bigr] \\
    &\times \Bigl[\delta(x_1'-d/2)-\delta(x_1'+d/2)\Bigr] \delta(t_1-t_1') \hat{\sigma}_x.
\end{split}\end{equation}

The Dyson equations for the matrix Green's function $\check{G}$ then read
\begin{equation}\begin{split}\label{eq:Dyson1}
    &\check{G}(x,t;x',t') = \check{G}_0(x,t;x',t')\\& + 
    \int dx_1 dt_1 dx_1' dt_1'\;
    \check{G}(x,t;x_1,t_1) \check{\Sigma}(x_1,t_1;x_1',t_1') \check{G}_0(x_1',t_1';x',t'),
\end{split}\end{equation}
and similarly
\begin{equation}\begin{split}\label{eq:Dyson2}
   & \check{G}(x,t;x',t') = \check{G}_0(x,t;x',t')\\& + 
    \int dx_1 dt_1 dx_1' dt_1'\;
    \check{G}_0(x,t;x_1,t_1) \check{\Sigma}(x_1,t_1;x_1',t_1') \check{G}(x_1',t_1';x',t').
\end{split}\end{equation}

We focus on Eq.~\eqref{eq:Dyson1} and consider $G^r$ only, since all transport quantities will be expressed in terms of it. The equation for the retarded Green's function reads
\begin{equation}\begin{split}
    &G^r(x,t;x',t') = G_0^r(x,t;x',t') \\
    &- i \frac{c_E}{2\pi d} \int dt_1 \;\Bigl[G^r(x,t;d/2,t_1) - G^r(x,t;-d/2,t_1)\Bigr] \\
    &\times \Bigl[G_0^r(d/2,t_1;x',t') - G_0^r(-d/2,t_1;x',t')\Bigr].
\end{split}\end{equation}

Using Eq.~\eqref{eq:bareretarded}, one can show that the last factor vanishes for $x' \geq d/2$, yielding $G^r = G_0^r$ in that domain. Similarly, from Eq.~\eqref{eq:Dyson2}, one finds $G^r = G_0^r$ for $x \leq -d/2$. Since we are not interested about what happens inside the junction, we focus on the domain $x \geq d/2$ and $x' \leq -d/2$, where the retarded Green's function takes the form
\begin{equation}\begin{split}\label{eqGr}
   & G^r(x,t;x',t') = i 2 \pi F(t - t' - x + x' + d)\\& - i \pi \theta(t - t') \sgn(t - t' - x + x' + d),
\end{split}\end{equation}
with the intermediate function $F$ defined as
\begin{equation}
    F(t) \equiv \frac{1}{2} \theta(t) + \frac{G^r(d/2,t;-d/2,0)}{i 2 \pi},\label{eq:F}
\end{equation}
and obeying the integral equation
\begin{equation}\label{F_int}
    F(t) = \theta(t) - \theta(t-d) - \frac{c_E}{d} \int_{t-d}^{t} dt_1 \; F(t_1).
\end{equation}
One can verify that $F$ is real and vanishes for negative times. Additionally, the advanced Green's function $G^a$ can be obtained from $G^r$ by the following general relation
\begin{equation}
    G^a(x,t;x',t')=G^r(x',t';x,t).
\end{equation}
The Keldysh component $G^K$ can be obtained by computing the corresponding projection of the Dyson equation in Eq.~(\ref{eq:Dyson1}), such that
\begin{equation}\begin{split}
    &G^K(x,t;x',t') = G_0^K(x,t;x',t') \\
    &- i \frac{c_E}{2\pi d} \int dt_1 \;\Bigl[G^K(x,t;d/2,t_1) - G^K(x,t;-d/2,t_1)\Bigr] \\
    &\times \Bigl[G_0^a(d/2,t_1;x',t') - G_0^a(-d/2,t_1;x',t')\Bigr] \\
    &- i \frac{c_E}{2\pi d} \int dt_1 \;\Bigl[G^r(x,t;d/2,t_1) - G^r(x,t;-d/2,t_1)\Bigr] \\
    &\times \Bigl[G_0^K(d/2,t_1;x',t') - G_0^K(-d/2,t_1;x',t')\Bigr].
\end{split}\end{equation}
For later calculations, we will only the need the Keldysh component at positions $x=-x'=d/2$. This allows us to simplify the above expression, since, according to the bare advanced Green's function in Eq.~\eqref{eq:bareadvanced}, one has 
\begin{equation}
    G_0^a\left(d/2,t_1;-d/2,t'\right) - G_0^a\left(-d/2,t_1;-d/2,t'\right) = 0.
\end{equation}
removing the first integral. The second one rewrites thanks to the previous discussion on $G^r$
\begin{equation}\begin{split}\label{eq:DysonGK<>}
    &G^K(d/2,t;-d/2,t') = 2\mathcal{G}_0(t'-t+d) \\
    &- 2\frac{c_E}{d} \int dt_1 \;F(t-t_1)
    \Bigl[\mathcal{G}_0(t'-t_1+d) - \mathcal{G}_0(t'-t_1)\Bigr].
\end{split}\end{equation}

\section{Current \label{sec:Current}}
We start by examining how the system responds in time when a state containing $M$ anyons is introduced by computing the time-dependent current. The details of the injection mechanism are deliberately left open at this stage, so that the discussion captures the essential physics without relying on a specific model.

The edge current is defined by $\hat{I}(x,t)=\rho(x,t)$. In the order to take into account the effects of tunneling at the junction, we use Keldysh perturbation theory in the Hamiltonian $H_T$ \cite{Martin05}. The average value of the current operator becomes
\begin{equation}
    I(x,t)\equiv\langle in|T_K\Bigl[\hat{I}(x,t^\xi)\;e^{-i\sum_{\eta=\pm}\eta\int dt_1\;H_T(t_1^\eta)}\Bigr]|in\rangle
\end{equation}
For consistency, we introduced a Keldysh index $\xi$ in the expression for the current $I(x,t)$: nevertheless, the latter will not play any role in the remainder of the calculation. In the perturbative limit, we can expand the result in orders in $H_T$. We firstly consider the zero-th order time-dependent current
\begin{equation}\begin{split}\label{I0}
    &I^{(0)}(x,t)=\langle in|T_K\Bigl[\hat{I}(x,t^\xi)\Bigr]|in\rangle
    \\&=-e\frac{\sqrt{\nu}}{2\pi}\partial_x\langle0|T_K\Bigl[
    \phi(x,t^\xi)e^{-i\sqrt{\nu}\sum_n\sum_\zeta\zeta\phi(-\mathcal{T},(-\mathcal{T}+\tau_n)^\zeta)}
    \Bigr]|0\rangle.
\end{split}\end{equation}

\begin{figure}
    \centering
    \includegraphics[scale=1]{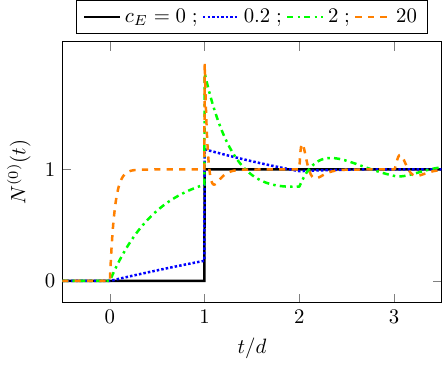}
    \caption{Equation (\ref{eqN}) with a unique injection at $\tau_1=d-D$. $N(t)$ expresses the amount of quasi-particles having reached the detector at time $t$. At zeroth order in tunneling, the only relevant parameter is the charging energy. At $c_E=0$, the injected quasi-particle reaches the detector at once at time $t=d$, corresponding to the delay of traveling the loop. For nonzero capacitive coupling, some part of the signal is advanced. When $c_E$ goes to infinity, the quasi-particle reaches the detector at once at $t=0$. Everything happens like if the capacitive coupling had allowed to jump the loop.
    }
    \label{figQ}
\end{figure}

The quantum averaging in Eq.~\eqref{I0} can be expressed in term of two points Green's functions using the bosonic identity
\begin{equation}
    \langle0|T_K\Bigl(e^{\sum_iq_i\phi_i}\Bigr)|0\rangle=e^{\frac{1}{2}\langle0|T_K\left[\left(\sum_iq_i\phi_i\right)^2\right]|0\rangle}
\end{equation}
that can be expanded into
\begin{equation}\begin{split}
    \langle0|T_K\left(\phi_ze^{\sum_iq_i\phi_i}\right)|0\rangle=&\langle0|T_K\left(\phi_z\sum_iq_i\phi_i\right)|0\rangle
    \\&\times e^{\frac{1}{2}\langle0|T_K\left[\left(\sum_iq_i\phi_i\right)^2\right]|0\rangle}
\end{split}\label{eq:identity}.\end{equation}
By using the above identity, the average of the current operator can be recast as
\begin{equation}\begin{split}\label{eq24}
    I^{(0)}&(x,t)=i\frac{e\nu}{2\pi}\partial_x\sum_n\sum_\zeta\zeta
    \\&\langle0|T_K\left\{
    \phi\left(x,t^\xi\right)\phi\left(-\mathcal{T},\left(-\mathcal{T}+\tau_n\right)^\zeta\right)
    \right\}|0\rangle
    \\&e^{-\frac{\nu}{2}\langle 0|T_K\left\{\left[\sum_{n'}\sum_{\zeta'}\zeta'\phi\left(-\mathcal{T},\left(-\mathcal{T}+\tau_{n'}\right)^{\zeta'}\right)\right]^2\right\}|0\rangle}
\end{split}\end{equation}
The argument of the exponential function is proportional to 
\begin{equation}
\sum_{\zeta'=\pm}\sum_{\zeta''=\pm}\zeta'\zeta'' G^{\zeta'\zeta''}\left(-\mathcal{T},-\mathcal{T}-\tau_{n'};-\mathcal{T},-\mathcal{T}-\tau_{m'}\right),
\end{equation}
which vanishes due to the identity \eqref{eq:UsefulIdentity2}. Therefore, the bare current becomes
\begin{equation}
    I^{(0)}(x,t)=i\frac{e\nu}{2\pi}\partial_x\sum_n\sum_\zeta\zeta G^{\xi\zeta}(x,t;-\mathcal{T}+\tau_n)
\end{equation}
Using the identity (\ref{eq:UsefulIdentity1}) the sum over $\zeta$ gives
\begin{equation}\begin{split}
    &I^{(0)}(x,t)=i\frac{e\nu}{2\pi}\partial_x\sum_nG^r(x,t;-\mathcal{T},-\mathcal{T}+\tau_n)
\end{split}\end{equation}
Finally, we compute the current at the detector position $x=D$ using Eq.~\eqref{eqGr}
\begin{equation}\begin{split}\label{eq:ZerothOrderCurrent}
    I^{(0)}(D,t)=-e\nu\sum_n&\delta(t-\tau_n-D+d)
    \\&-F'(t-\tau_n-D+d)
\end{split}\end{equation}
This bare current can be interpreted as a stream of delta-like quasi-particles propagating chirally along the edge state, modified by an additional chiral contribution caused by capacitive effects. This marks already a first difference compared with the result from Refs.~\cite{Ronetti25,Ronetti25b}. To visualize this effect, we plot in Fig.~\ref{figQ} the integral over time, up to a time $t$, of the bare current $I^{(0)}$, corresponding to the number of quasi-particles having reached the detector at a certain instant $t$.
\begin{equation}\begin{split}\label{eqN}
    &N^{(0)}(t)\equiv\frac{1}{-e\nu}\int_{-\infty}^tdt'\;I^{(0)}(D,t')
    \\=&\sum_n\left[\theta(t-\tau_n-D+d)-F(t-\tau_n-D+d)\right].
\end{split}\end{equation}
While this sets the stage for our calculation and provides some initial insights into the effect of capacitive coupling, the bare current remains unaffected by braiding and therefore cannot reveal information about the statistical angle $\lambda$. To probe this effect, we must include the contribution of the tunneling Hamiltonian $H_T$, which accounts for time-domain braiding processes between incoming quasi-particles and those generated at the junction. To this end, we compute the current to first order in tunneling using the Keldysh formalism
\begin{equation}
    I^{(1)}(x,t)=-i\langle in|T_K\Bigl[\hat{I}(x,t^\xi)\sum_\eta\eta\int dt_1\;H_T(t_1^\eta)\Bigr]|in\rangle.
\end{equation}
Its bosonized expression is 
\begin{equation}\begin{split}
    &I^{(1)}(x,t)=ie\frac{\sqrt{\nu}}{2\pi}\partial_x\sum_\eta\eta\int dt_1\;\frac{\Gamma}{2\pi a}\sum_\epsilon e^{-i\epsilon\kappa}
    \\&\langle0|T_K\Bigl\{\phi(x,t^\xi)e^{i\epsilon\sqrt{\nu}\left[\phi(d/2,t_1^\eta)-\phi(-d/2,t_1^\eta)\right]}
    \\&e^{-i\sqrt{\nu}\sum_n\sum_\zeta\zeta\phi(-\mathcal{T},(-\mathcal{T}+\tau_n)^\zeta)}\Bigr\}|0\rangle.
\end{split}\end{equation}
The above expression can be readily recast in terms of more elementary Keldysh averages by resorting to the identity in Eq.~\eqref{eq:identity}
\begin{equation}\begin{split}\label{eqIH}
&I^{(1)}(x,t)=ie\frac{\sqrt{\nu}}{2\pi}\partial_x\sum_\eta\eta\int dt_1\;\frac{\Gamma}{2\pi a}\sum_\epsilon e^{-i\epsilon\kappa}
     \\&\times e^{\epsilon\nu\langle0|T_K\left\{\bigl[\phi(d/2,t_1^\eta)-\phi(-d/2,t_1^\eta)\bigr]\bigl[\sum_n\sum_\zeta\zeta\phi(-\mathcal{T},(-\mathcal{T}+\tau_n)^\zeta)\bigr]\right\}|0\rangle}
    \\&\times e^{-\frac{\nu}{2}\langle0|T_K\Bigl\{\bigl[\phi(d/2,t_1^\eta)-\phi(-d/2,t_1^\eta)\bigr]^2\Bigr\}|0\rangle}
    \\&\times\langle0|T_K\Bigl\{\phi(x,t^\xi)\Bigl[i\epsilon\sqrt{\nu}\phi(d/2,t_1^\eta)-i\epsilon\sqrt{\nu}\phi(-d/2,t_1^\eta)
    \\&\qquad\qquad\qquad-i\sqrt{\nu}\sum_m\sum_{\zeta'}\zeta'\phi(-\mathcal{T},(-\mathcal{T}+\tau_m)^{\zeta'})\Bigr]
    \Bigr\}|0\rangle.
\end{split}\end{equation}
Some simplifications can be done in the previous expression. We use Eq. (\ref{eq:UsefulIdentity1}) to recast the exponent in the second line. For the exponent of the third line, the Keldysh prescription implies that
\begin{equation}
T_K\bigl(A(t_1^\eta)B(t_1^\eta)\bigr)=\bigl[A(t_1)B(t_1)+B(t_1)A(t_1)\bigr]/2,
\end{equation}
providing
\begin{equation}\begin{split}\label{eq:preGKd}
    \langle0|T_K\Bigl\{\bigl[\phi(d/2,t_1^\eta)-\phi(-d/2,t_1^\eta)&\bigr]^2\Bigr\}|0\rangle
    \\=-G^K&(d/2,t_1;-d/2,t_1)
\end{split}\end{equation}
In the following, we will use the shortened notation $\mathcal{G}(d)$. Its expression can be computed thanks to Eq.~(\ref{eq:DysonGK<>}).
\begin{equation}\begin{split}\label{eq:GKd}
    &\mathcal{G}(d)\equiv\frac{1}{2}G^K(d/2,t,-d/2,t)
    \\=&\mathcal{G}_0(d)-\frac{c_E}{d} \int dt_2 \;\bigl[\mathcal{G}_0(t_2+d) - \mathcal{G}_0(t_2)\bigr]F(t_2).
\end{split}\end{equation}
The sum on $\eta$ is reported on the last two lines. Since the term in the very last line is independent of $\eta$ the sum over this variable will give zero. For the fourth line, we use Eq.~\eqref{eq:UsefulIdentity1}. As a consequence, one obtains
\begin{equation}\begin{split}\label{eq34}
    &I^{(1)}(x,t)=\frac{-e\nu\Gamma}{4\pi^2a}e^{\nu\mathcal{G}(d)}
    \int dt_1\sum_\epsilon\epsilon e^{-i\epsilon\kappa}
    \\&\partial_x\bigl[G^r(x,t;d/2,t_1)-G^r(x,t;-d/2,t_1)\bigr]
    \\&e^{\epsilon\nu\sum_nG^r(d/2,t_1;-\mathcal{T},-\mathcal{T}+\tau_n)-G^r(-d/2,t_1;-\mathcal{T},-\mathcal{T}+\tau_n)}
\end{split}\end{equation}
We recall that for $x<-d/2$, the retarded Green's function $G^r$ assumes its bare value given in Eq.~\eqref{eq:bareretarded}. Therefore, consistently with the causality, the time-dependent current vanishes before the junction. At detector position $x=D>d/2$, we can insert the result for the retarded Green's function computed in presence of the capacitive coupling appearing in Eq.~\eqref{eqGr},
\begin{equation}\begin{split}\label{eq42}
    I^{(1)}(D,t)=&\frac{e\nu\Gamma}{\pi a}e^{\nu\mathcal{G}(d)}\int dt_1\;
    F'(t-t_1-D+d/2)
    \\&\sin\Bigl[2\pi\nu\sum_nF(t_1+d/2-\tau_n)-\kappa\Bigr].
\end{split}\end{equation}

In the theory, $\nu$ can correspond either to the fractional charge, scaling dimension and statistical angle. We will distinguish between these different roles by hand using the following rules : When $\nu$ is next to $e$, it corresponds to the fractional charge $\nu_c$. Else, it correspond to the scaling dimension $\nu_\delta$ in front of $\{\phi,\phi\}/G^K/\mathcal{G}$ and to the statistical angle $\nu_\lambda$ for $[\phi,\phi]/G^a/G^r/F$. These rules ensure that for $\Theta=c_E=0$ :
\begin{equation}\begin{split}
    \langle\psi(x,t)\psi^\dagger(x',t')\rangle=\frac{1}{2\pi a}\left|\frac{ia}{ia+t'-t-x'+x}\right|^{\nu_\delta}&
    \\\times\; e^{sgn(t'-t-x'+x)\,i\frac{1}{2}\pi\nu_\lambda}&.
\end{split}\end{equation}
$\nu_\delta$ is related to the decay of correlations with time while $\nu_\lambda$ gives the braiding phase. In Eq.~\eqref{eq42}, one gets
\begin{equation}\begin{split}\label{I_F}
    I^{(1)}(D,t)=&I^{(1)}_0\int dt_1\;
    F'(t-t_1-D+d/2)
    \\&\sin\Bigl[2\pi\nu_\lambda\sum_nF(t_1+d/2-\tau_n)-\kappa\Bigr].
\end{split}\end{equation}
\begin{figure}
    \centering
    \includegraphics[scale=1]{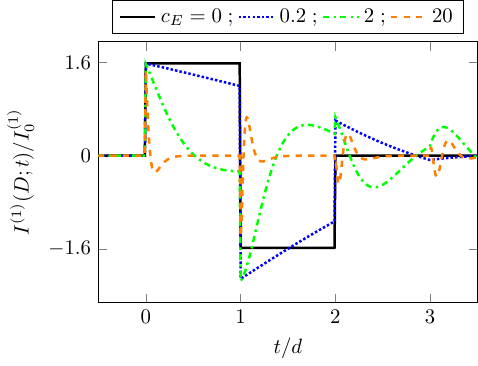}
    \caption{First order current obtained in Eq.~\eqref{I_F}, with a unique injection at $\tau_1=d-D$ and divided by the prefactor $\frac{e\nu\Gamma}{\pi a}e^{\nu\mathcal{G}(d)}$. Different values of $c_E$ are used. Other parameters used are $\nu_\lambda=1/3$ and $\kappa=\pi/5$ so $\sin(2\pi\nu_\lambda-\kappa)-\sin(-\kappa)\simeq1.6$. For $c_E=0$, we find back the result of \cite{Ronetti25}. As $c_E$ increases, the signal takes longer time to vanish but peaks get thinner. Thus, there is almost no first-order current left at $c_E=20$.
    }
    \label{figI}
\end{figure}
As in \cite{Ronetti25b}, a clear separation arises between the prefactor
\begin{equation}
I^{(1)}_0 \equiv \frac{e\nu_c\Gamma}{\pi a}e^{\nu_\delta\mathcal{G}(d)}
\end{equation}
and the integral part containing only $\nu_\lambda$. This last one is represented in fig \ref{figI}. 

\section{Cross correlations \label{sec:Noise}}

As a consequence of the conservation of the current before and after the loop, any nontrivial order in perturbation is identically 0 in average. At order 1, one can get convinced of that by integrating $F'(t-...)$ over all times in equation (\ref{I_F}). 
Time resolved measurement being out of reach for current detectors, one has to find an other observable.

In order to assess the validity of the protocol introduced in Refs.~\cite{Ronetti25,Ronetti25b} in presence of the capacacitive coupling, we compute the finite frequency cross-correlation noise between reservoirs $D$ and $D_L$ {(see Fig.~\ref{schOm} for a sketch of the setup)}. This goal will be achieved by using the non-equilibrium bosonization formalism as in Ref.~\cite{Rosenow16}, taking into account the presence of both edges. 

The finite-frequency cross-correlation noise is defined as~\cite{Chevallier10}
\begin{equation}\begin{split}
    &\mathcal{S}=\int d(t-t_2)\;\cos\bigl(\omega(t-t_2)\bigr)\Bigl\langle\delta\hat{I}_R(D,t)\delta\hat{I}_L(D_L,t_2)\Bigr\rangle_{th,st}
    \label{eq:FinitFreqCorr}
\end{split}\end{equation}
where we introduced the current fluctuations
\begin{equation}\begin{split}
\delta \hat{I}_{R/L}(x,t) = \hat{I}_{R/L}(x,t) 
 - \left\langle\hat{I}_{R/L}(x,t)\right\rangle_{th,st} \, .
\end{split}\end{equation}
The average in Eq.~\eqref{eq:FinitFreqCorr} is at the same time a thermal average and a Poissonian average. These two averages are independent. Since thermal averaging can be performed independently on each edge, the cross-correlation noise becomes
\begin{equation}\begin{split}
    &\mathcal{S}=\int d(t-t_2)\;\cos\bigl[\omega(t-t_2)\bigr]
    \\&\biggl[\Bigl\langle\bigl\langle\hat{I}_R(D,t)\bigr\rangle_{th}\bigl\langle\hat{I}_L(D_L,t_2)\bigr\rangle_{th}\Bigr\rangle_{st}
    \\&-\Bigl\langle\bigl\langle\hat{I}_R(D,t)\bigr\rangle_{th}\Bigr\rangle_{st}\Bigl\langle\bigl\langle\hat{I}_L(D_L,t_2)\bigr\rangle_{th}\Bigr\rangle_{st}\biggr]
\end{split}\end{equation}
$\langle\hat{I}_R\rangle_{th}$ is exactly the previously computed quantity $I^{(0,1)}$, depending the order in tunneling. $\langle\hat{I}_L\rangle_{th}$ can be accessed easily thanks to the conservation of the current at the injection QPC :
\begin{equation}\label{cons_courant_CPS}
    I_{drive}=\langle\hat{I}_L(-\mathcal{T},t')\rangle_{th}+\langle\hat{I}_R(-\mathcal{T},t')\rangle_{th}
\end{equation}
Using chirality of the both edge, the above relation can be turned into
\begin{equation}
    I_{drive}=\langle\hat{I}_L(D_L,t_2)\rangle_{th}+\left[I^{(0)}(D',t_2)\right]_{c_E=0}
\end{equation}
We wrote $I^{(0)}$ and $c_E=0$ to keep in mind that the current at $D'$ has been obtained only by translation of coordinates and thus does not account any effect of $H_T$ or $H_E$. The position $D'=-2\mathcal{T}-D_L$ is obtained by symmetry of $D_L$ by the injection QPC (see fig~\ref{schOm}).

Using the fact that $I_{drive}$ is constant regarding statistical averaging, Eq.~\eqref{eq:FinitFreqCorr} rewrites ($j=0,1$)
\begin{equation}\begin{split}
    &\mathcal{S}^{(j)}=-\int d(t-t_2)\;\cos\bigl[\omega(t-t_2)\bigr]
    \\&\biggl\{\Bigl\langle I^{(j)}(D,t)\;\left[I^{(0)}(D',t_2)\right]_{c_E=0}\Bigr\rangle_{st}
    \\&-\Bigl\langle I^{(j)}(D,t)\Bigr\rangle_{st}\Bigl\langle \left[I^{(0)}(D',t_2)\right]_{c_E=0}\Bigr\rangle_{st}
    \biggr\}.
\end{split}\end{equation}
In the remainder of this section, we will compute the statistical average of these two cross-correlations contributions. In order to do that, we should consider explicitly each separate case for $j=0,1$. We start by writing the explicit expression of $\mathcal{S}^{(0)}$
\begin{widetext}
\begin{equation}\begin{split}
    \mathcal{S}^{(0)}=-e^2\nu_c^2\int&d(t-t_2)\;\cos\bigl[\omega(t-t_2)\bigr]
    \\\times\biggl(&\Bigl\langle\sum_n\Bigl\{\delta(t-D-\tau_n)+\frac{c_E}{d}\bigl[F(t-D-\tau_n+d)-F(t-D-\tau_n)\bigr]\Bigr\}\sum_m\delta(t_2-D'-\tau_m)\Bigr\rangle_{st}
    \\-&\Bigl\langle\sum_n\Bigl\{\delta(t-D-\tau_n)+\frac{c_E}{d}\bigl[F(t-D-\tau_n+d)-F(t-D-\tau_n)\bigr]\Bigr\}\Bigr\rangle_{st}\Bigl\langle\sum_m\delta(t_2-D'-\tau_m)\Bigr\rangle_{st}
    \biggr).
\end{split}\end{equation}
\end{widetext}
The statistical average of the above cross-correlation can be computed using the results in Appendix~\ref{app:StatisticalAverage}. In particular, by exploiting the relation in Eq.~\eqref{eq:StatisticalAverage2}, one arrives to
\begin{equation}\begin{split}
    &\mathcal{S}^{(0)}=-e^2\nu_c^2\gamma\int dt\;\cos(\omega t)
    \Bigl\{\delta(t+D'-D)
    \\&+\frac{c_E}{d}\bigl[F(t+D'-D+d)-F(t+D'-D)\bigr]\Bigr\}.
\end{split}\end{equation}
In the above expression, the quantity of interest, namely the statistical parameter $\pi\lambda$, does not explicitly appear, as this zeroth-order contribution neglects the effect of tunneling at the junction, where time-domain braiding processes take place. To capture these effects and access $\pi\lambda$, it is therefore necessary to consider the first-order contribution in the tunneling Hamiltonian. The latter reads
\begin{equation}\begin{split}
    &\mathcal{S}^{(1)}=\frac{e^2\nu_c^2\Gamma}{\pi a}e^{\nu_\delta\mathcal{G}(d)}\int dt\;\cos\bigl(\omega t\bigr)
    \\&\int dt_1
    F'(t+t_2-t_1+D'-D+d)
    \\&\Im\biggl\{e^{-i\kappa}\biggl[\Bigl\langle e^{i2\pi\nu_\lambda\sum_nF(t_1-\tau_n)}\sum_m\delta(t_2-\tau_m)\Bigr\rangle_{st}
    \\&-\Bigl\langle e^{i2\pi\nu_\lambda\sum_nF(t_1-\tau_n)}\Bigr\rangle_{st}\Bigl\langle\sum_m\delta(t_2-\tau_m)\Bigr\rangle_{st}
    \biggl]\biggr\}
\end{split}\end{equation}

To evaluate the statistical average of the cross-correlation, we make use of the identities derived in Appendix~\ref{app:StatisticalAverage}. Specifically, by applying the result of Eqs.~\eqref{eq:StatisticalAverage3} and~\eqref{eq:StatisticalAverage4}, one can find the final expression of the finite-frequency cross-correlation in the following compact form
\begin{equation}\begin{split}\label{eqS1}
    &\mathcal{S}^{(1)}=-\gamma\frac{e^2\nu_c^2\Gamma}{\pi a}e^{\nu_\delta\mathcal{G}(d)}
    \Im\biggl[e^{-i\kappa}e^{-\gamma\int dt'\;1-e^{i2\pi\nu_\lambda F(t')}}
    \mathcal{F}\biggr],
\end{split}\end{equation}
where we introduced the function
\begin{equation}\begin{split}
    \mathcal{F}\equiv&\int dt\;\cos\bigl(\omega t\bigr)\int dt"F'(t-t"+D'-D+d)
    \\&\times \left[1-e^{i2\pi\nu_\lambda F(t")}\right].\label{eq:mathcalF}
\end{split}\end{equation}
In order to make contact with the previous results found in Refs.~\cite{Ronetti25,Ronetti25b}, we compute the expression in the absence of the capacitive coupling, i.e. for $c_E=0$. The function appearing in Eq.~\eqref{eq:mathcalF} becomes
\begin{equation}\begin{split}
    \mathcal{F}_{c_E=0}=-ie^{i\pi\nu_\lambda}\sin(\pi\nu_\lambda)\frac{8}{\omega}\sin\bigl[\omega(D-D')\bigr]\sin^2(\omega d/2)
\end{split}\end{equation}
and, therefore, we find
\begin{equation}\begin{split}
    &\mathcal{S}^{(1)}_{c_E=0}=4\frac{e^2\nu_c^2\Gamma}{\pi a}e^{\nu_\delta\mathcal{G}_0(d)}\sin\bigl[\omega(D-D')\bigr]\frac{\sin^2(\omega d/2)}{\omega d/2}
    \times
    \\&\sin(\pi\nu_\lambda)\gamma de^{-\gamma d[1-\cos(2\pi\nu_\lambda)]}\cos\bigl[\gamma d\sin(2\pi\nu_\lambda)+\pi\nu_\lambda-\kappa\bigr].
\end{split}\end{equation}

By varying the product $\gamma d$, one observes a sequence of regularly spaced zeros in the cross-correlation, originating from the cosine term 
\(\cos\bigl(\gamma d \sin(2\pi \nu_\lambda) + \dots\bigr)\). The spacing between consecutive zeros is determined by the condition that the argument of the cosine changes by $\pi$, yielding
\begin{equation}
    \Delta (\gamma d) = \frac{\pi}{\sin(2\pi \nu_\lambda)}.
\end{equation}
This periodic pattern directly reflects the braiding phase, thus allowing to extract the statistical angle without the need of knowing the scaling dimension.~\cite{Ronetti25,Ronetti25b}.

\begin{figure}[h!]
    \centering
    \includegraphics[scale=1]{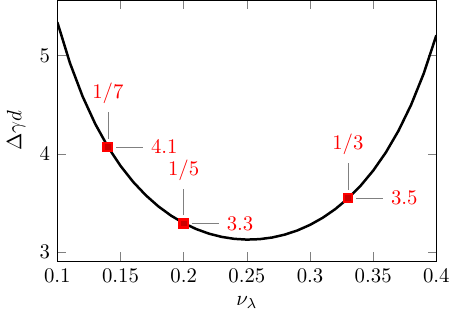}
    \includegraphics[scale=1]{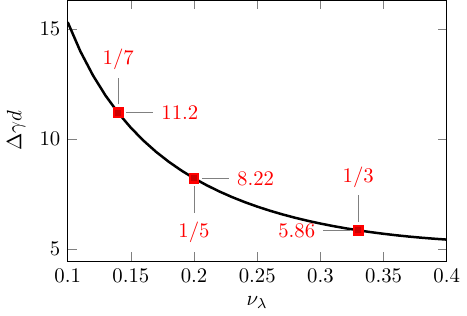}
    \caption{Spacing between zeros without charging energy (top) and with $c_E=2$ (bottom). The curves correspond to the prediction $\pi d/\int dt\, \sin(2\pi \nu_\lambda F(t))$ as a function of $\nu_\lambda$. Knowing $c_E$, one can measure $\Delta (\gamma d)$ and infer the corresponding value of $\nu_\lambda$. For low capacitance, multiple statistical angles can correspond to the same $\Delta (\gamma d)$.}
    \label{S1}
\end{figure}

In the more general case described by Eq.~\eqref{eqS1}, the cross-correlation is proportional to 
\begin{equation}
\sin\left\{\gamma \int dt' \, \sin\bigr[2\pi \nu_\lambda F(t')\bigl] + \arg(\mathcal{F}) - \kappa \right\},
\end{equation}
where the function $F(t)$ encodes the time-dependent effects of the tunneling events and the capacitive coupling.  Here, the spacing between zeros is modified by the presence of the capacitance, and is given by
\begin{equation}
    \Delta (\gamma d) = \frac{\pi d}{\int dt' \, \sin\bigl[2\pi \nu_\lambda F(t')\bigr]}~.
\end{equation}
This relation is illustrated in Fig.~\ref{S1}, which compares the spacing between zeros in the absence and in the presence of a finite charging energy $c_E$. Here, we show how the spacing between zeros evolves as the charging energy is increased, highlighting the impact of the capacitive interaction on the signatures of the braiding phase. In this situation, the knowledge of the charging energy is crucial to extract the statistical angle $\pi\lambda$. In the next section, we present a protocol to overcome this issue.

Finally, a notable difference compared to the case $c_E = 0$ is that, previously, the cross-correlations vanished for $\nu_\lambda = 1$; this is no longer the case when a finite charging energy is present.

%\begin{figure}
 %   \centering
  %  \includegraphics[width=0.99\linewidth]{lambda1.pdf}
   % \caption{$\frac{1}{d}\int dt'\;\sin(2\pi\nu_\lambda F(t'))$ at $\nu_\lambda=1$ plotted as a function of $c_E$. There is not much to say.}
   % \label{fig_lam1}
%\end{figure}

\section{Gate voltage \label{sec:Gate}}
In order to extract the statistical angle $\pi\lambda$ directly, it is necessary to obtain an independent measurement of the charging energy $c_E$. To achieve this, the experimental setup can be completed by a metallic gate capacitively coupled to the loop. Applying a DC voltage $V_G$ on the gate modifies the total charge seen by the capacitor according to
\begin{equation}
    Q = -e\frac{\sqrt{\nu_c}}{2\pi}\Bigl[\phi(d/2)-\phi(-d/2)\Bigr] + C V_G,
\end{equation}
where the first term accounts for the edge excitations and the second for the gate-induced charge.  

The charging energy now acquires additional contributions
\begin{equation}
    H_E \rightarrow H_E + H_G + \frac{C}{2} V_G^2,
\end{equation}
where the last term is a constant that can be ignored, and the cross term
\begin{equation}
    H_G = -e\frac{\sqrt{\nu_c}}{2\pi} V_G \Bigl[\phi(d/2)-\phi(-d/2)\Bigr],
\end{equation}
represents the capacitive coupling between the edge excitations and the gate.  

This coupling can be incorporated naturally in the Keldysh formalism together with the tunneling Hamiltonian $H_T$, yielding the gated current
\begin{equation}
    I_{\rm gated}(D,t) = \langle in|T_K\Bigl[\hat{I}(D,t^\xi)\, e^{-i\sum_\eta \eta \int dt_1 \,(H_T + H_G)(t_1^\eta)}\Bigr]|in\rangle.
\end{equation}

Since $H_T$ and $H_G$ commute, we can compute the current perturbatively to first order in $H_T$ while keeping all orders in $H_G$
\begin{equation}\begin{split}\label{eq:WickGate}
    I^{(1)}_{\rm gated}(D,t) &= \langle in| T_K \Bigl[ \hat{I}(D,t^\xi) \, \times \\
     -i \sum_\eta \eta \int dt_1&\; H_T(t_1^\eta) \, e^{-i \sum_{\eta'} \eta' \int dt_1' H_G(t_1'^{\eta'})} \Bigr] |in\rangle.
\end{split}\end{equation}
The details of this calculation are presented in Appendix~\ref{app:Gate}. The effect of the gate is simply to add a voltage-dependent phase shift to the first-order current in Eq.~\eqref{I_F}, such that 
\begin{equation}
\varphi_{\rm ungated} = -\kappa \quad\rightarrow\quad \varphi_{\rm gated} = -\kappa+\frac{e \nu_c V_G d}{1 + c_E}.
\end{equation}
The same modification applies to the first-order cross correlations, which now read
\begin{equation}\begin{split}
    &\mathcal{S}^{(1)}_{\rm gated}(\gamma,\kappa,V_G) = -\gamma \frac{e^2 \nu_c^2 \Gamma}{\pi a} e^{\nu_\delta \mathcal{G}(d)} \\
    & \quad \times \Im \Biggl[ e^{-i\kappa + i e \nu_c V_G d/(1+c_E)} 
    \, e^{-\gamma \int dt' \,(1 - e^{i 2 \pi \nu_\lambda F(t')})} \, \mathcal{F} \Biggr].
\end{split}\end{equation}

In this expression, the experimentalist has direct control over $V_G$, $\gamma$, and $\kappa$, where the latter includes the contribution of the Aharonov-Bohm phase due to the external magnetic flux threading the loop.  
Consider the following situation: starting from a zero of $\mathcal{S}^{(1)}_{\rm gated}$, one varies the parameters slowly while remaining on the zero. This condition can be written as
\begin{equation}\label{eq:phase_zero}
    -\Delta \kappa + \Delta V_G \frac{e \nu_c d}{1 + c_E} + \Delta \gamma \int dt' \, \sin\bigl[2 \pi \nu_\lambda F(t')\bigr] = 0.
\end{equation}

If we keep $\gamma$ constant, this relation provides a direct measurement of the charging energy:
\begin{equation}
    c_E = -1 + e \nu_c d \left. \frac{\Delta V_G}{\Delta \kappa} \right|_{\Delta \gamma = 0}.
\end{equation}
Alternatively, by controlling all three parameters, Eq.~\eqref{eq:phase_zero} can be used to determine $\int dt' \, \sin\bigl[2\pi\nu_\lambda F(t')\bigr]$ and ultimately extract the statistical angle $\nu_\lambda$.

\section{Conclusion \label{sec:Conclusions}}

In this work, we have analyzed the single-edge interferometer recently proposed in Refs.~\cite{Ronetti25,Ronetti25b}, focusing on the influence of capacitive coupling along the loop. This geometry, in which a quantum point contact (QPC) defines a loop along a single edge, allows tunneling to be controlled via a gate voltage and provides direct access to the universal statistical angle $\pi\lambda$ in the weak backscattering regime. In the absence of capacitance, first-order perturbation theory yields a clear factorization of transport quantities into a non-universal prefactor and a braiding-dependent term, enabling a straightforward extraction of $\lambda$ from current cross-correlations.

Here, we have extended the analysis by including the effect of a finite charging energy, which modifies both the current and its fluctuations. Using a two-point Green's function formalism supplemented with Dyson’s equation, we have shown that although the statistical angle $\lambda$ continues to imprint on the cross-correlations, the observed signal now also depends explicitly on the loop capacitance. As a result, a direct measurement of $\lambda$ requires independent knowledge of the charging energy. We have proposed an experimentally feasible scheme to achieve this, by capacitively coupling a metallic gate to the junction and monitoring the corresponding voltage-dependent phase shifts.

Our results demonstrate that the extraction of the statistical angle in single-edge interferometers is robust, provided that capacitive effects are properly accounted for. This work thus bridges the gap between idealized theoretical models and realistic experimental implementations, highlighting the importance of controlling both tunneling and charging effects in mesoscopic interferometry.

Although our analysis has focused on Laughlin fractions, the approach can be readily extended to more general fractional quantum Hall states, including non-Abelian or other non-integer fillings \cite{Nayak08,Cano13}.

\acknowledgments{M. H. would like to thank Victor Bastidas for fruitful discussions at the very first stage of this work. This work was carried out in the framework of the project ``ANY-HALL" (ANR Grant No.
ANR-21-CE30-0064-03). It received support from the French government under the France 2030 investment plan, as part of the Initiative d’Excellence d’Aix-Marseille Université A*MIDEX. We acknowledge support from the institutes IPhU (AMX-19-IET008) and AMUtech (AMX-19-IET-01X). This work has benefited from State aid managed by the Agence Nationale de la Recherche under the France 2030 programme, reference ``ANR-22-PETQ-0012".
This French-Japanese collaboration is supported by the CNRS International Research Project ``Excitations in Correlated Electron Systems driven in the GigaHertz range" (ESEC). This work was supported by Grants-in-Aid for Scientific Research (Grant 
Nos. JP24H00827 and JP22H00112) and the Japan Science and Technology Agency (JST) ASPIRE Program No. JPMJAP2410.}

\appendix
\section{Statistical average of anyons \label{app:StatisticalAverage}}
In this Section, we compute the general form of statistical averages appearing in the main text. In our setup, anyons are emitted from a source quantum point contact (QPC), drawing inspiration the collider geometry~{\cite{Bartolomei20,Rosenow16,Han16,Lee22,Morel22,Mora22,Schiller23,Jonckheere23}}. This source QPC is placed upstream with respect to the loop junction and is tuned to the weak backscattering regime. Under this regime, it is well established that the tunneling of anyons occurs following a Poissonian distribution, when a constant DC bias $V_{DC}$ is applied across the source QPC. The distribution of the arrival times of anyon excitations is, therefore, an uniform distribution.

The treatment of statistical averaging is the following. The time axis is discretized with an arbitrary small interval $\Delta t$. At each step $n$, corresponding to a tunneling event, there is a possibility of emitting an anyon quasi-particle. These events are represented by a family of random variables $\{\hat{X_n}\}_{n\in\mathbf{Z}}$. The $\hat{X_n}$ are independent, identically distributed and follow a Bernoulli law of parameter $\gamma\Delta t$, $\gamma$ being the average current at the output of the source QPC. In the end of the calculation, we will consider the continuum limit $\Delta t \rightarrow 0$ and $n\rightarrow \infty$. We can schematize this change of representation as follows
\begin{equation}
    \sum_nf(\tau_n)\Rightarrow\sum_{n\in\mathbf{Z}}\hat{X}_nf(n\Delta t)
\end{equation}
On such sum, the averaging procedure consists in taking the average value of each $\hat{X_n}$
\begin{equation}\begin{split}
    \Bigl\langle\sum_{n\in\mathbf{Z}}\hat{X}_nf(n\Delta t)\Bigr\rangle_{st}&=\sum_{n\in\mathbf{Z}}\gamma\Delta t\;f(n\Delta t).
\end{split}\end{equation}
In the continuum limit, one obtains
\begin{equation}\begin{split}
    \Bigl\langle\sum_{n\in\mathbf{Z}}\hat{X}_nf(n\Delta t)\Bigr\rangle\rightarrow{}\gamma\int dt_3\;f(t_3).\label{eq:StatisticalAverage1}
\end{split}\end{equation}
It is also useful to compute the following cross-correlation
\begin{equation}\begin{split}
    &\Bigl\langle\sum_nf(\tau_n)\sum_mg(\tau_m)\Bigr\rangle_{st}-\Bigl\langle\sum_nf(\tau_n)\Bigr\rangle_{st}\Bigl\langle\sum_mg(\tau_m)\Bigr\rangle_{st}
    \\&=\Bigl\langle\sum_nf(\tau_n)g(\tau_n)\Bigr\rangle_{st}+\Bigl\langle\sum_nf(\tau_n)\Bigr\rangle_{st}\Bigl\langle\sum_{m\ne n}g(\tau_m)\Bigr\rangle_{st}\\&-\Bigl\langle\sum_nf(\tau_n)\Bigr\rangle_{st}\Bigl\langle\sum_mg(\tau_m)\Bigr\rangle_{st}.
\end{split}\end{equation}
We use the fact that the indices over which we are summing are an infinite number, such that 
\begin{equation}
\Bigl\langle\sum_{m\ne n}g(\tau_m)\Bigr\rangle_{st} = \Bigl\langle\sum_mg(\tau_m)\Bigr\rangle_{st},
\end{equation}
and one obtains
\begin{equation}\begin{split}
    &\Bigl\langle\sum_nf(\tau_n)\sum_mg(\tau_m)\Bigr\rangle_{st}-\Bigl\langle\sum_nf(\tau_n)\Bigr\rangle_{st}\Bigl\langle\sum_mg(\tau_m)\Bigr\rangle_{st}
    \\&=\Bigl\langle\sum_nf(\tau_n)g(\tau_n)\Bigr\rangle_{st}.
\end{split}\end{equation}
Therefore, by using Eq.~\eqref{eq:StatisticalAverage1}, the cross-correlation in the continuum limit gives
\begin{equation}\begin{split}
    &\Bigl\langle\sum_nf(\tau_n)\sum_mg(\tau_m)\Bigr\rangle_{st}-\Bigl\langle\sum_nf(\tau_n)\Bigr\rangle_{st}\Bigl\langle\sum_mg(\tau_m)\Bigr\rangle_{st}
    \\&\xrightarrow{}\gamma\int dt_3\;f(t_3)g(t_3)\label{eq:StatisticalAverage2}.
\end{split}\end{equation}

In order to address all the statistical averages appearing in the main text, we also need to consider an exponential of the form $e^{\sum_n f(\tau_n)}$. By using the formulation in terms of the Poisson variable $\hat{X}_n$, the exponential can be rewritten as
\begin{equation}
    e^{\sum_n f(\tau_n)}
    \;\Rightarrow\;
    e^{\sum_{n\in\mathbf{Z}} \hat{X}_n f(n\Delta t)}
    = \prod_{n\in\mathbf{Z}} e^{\hat{X}_n f(n\Delta t)}.
\end{equation}

The stochastic average can then be computed by exploiting the statistical independence of the random variables $\hat{X}_n$. For each time slice, the variable $\hat{X}_n$ takes the value $1$ with probability $\gamma \Delta t$ and $0$ otherwise, so that
\begin{equation}
\begin{split}
    \Bigl\langle \prod_{n\in\mathbf{Z}} e^{\hat{X}_n f(n\Delta t)} \Bigr\rangle_{st}
    &= \prod_{n\in\mathbf{Z}}
    \left( \gamma \Delta t \, e^{f(n\Delta t)} + (1 - \gamma \Delta t) \, \right).
\end{split}
\end{equation}

To proceed, we now take the continuous-time limit.  
Expanding the product to first order in $\Delta t$ and using
\[
\prod_n (1 + a_n \Delta t) \;\xrightarrow[]{}\; e^{\sum_n a_n \Delta t} \;\to\; e^{\int dt\, a(t)},
\]
we obtain the compact expression
\begin{equation}
    \Bigl\langle e^{\sum_n f(\tau_n)} \Bigr\rangle_{st}
    \xrightarrow[]{}
    e^{-\gamma \int dt_3 \, [1 - e^{f(t_3)}]}.\label{eq:StatisticalAverage3}
\end{equation}

\vspace{0.3cm}

A similar reasoning applies when an additional sum over functions of the event times appears inside the average.  
For instance, for an expression of the type $\bigl\langle e^{\sum_n f(\tau_n)} \sum_m g(\tau_m) \bigr\rangle_{st}$, the discrete representation yields
\begin{equation}
\begin{split}
    \Bigl\langle e^{\sum_n f(\tau_n)} \sum_m g(\tau_m) \Bigr\rangle_{st}
    &= \sum_m g(m\Delta t)\,
    \gamma \Delta t\, e^{f(m\Delta t)}\\&
    \times\prod_{n \neq m}
    \left( \gamma \Delta t \, e^{f(n\Delta t)} + (1 - \gamma \Delta t) \right).
\end{split}
\end{equation}

Taking again the continuous limit, one obtains
\begin{equation}
\begin{split}
    \Bigl\langle e^{\sum_n f(\tau_n)} \sum_m g(\tau_m) \Bigr\rangle_{st}
    \;\longrightarrow\;
    \gamma \int dt_3 \, g(t_3) e^{f(t_3)} &\\
    \times\,e^{-\gamma \int dt_4 \, [1 - e^{f(t_4)}]} &.
\label{eq:StatisticalAverage4}
\end{split}
\end{equation}

\section{Derivation of the gated current \label{app:Gate}}

In this appendix, we present the detailed derivation of the first-order current in the presence of a gate voltage, corresponding to Eq.~\eqref{eq:WickGate} in the main text. We aim to show explicitly how the gate-induced phase arises from the bosonic representation of the edge fields and how it connects to the charging energy $c_E$.

We start from the Keldysh expression at first order in $H_T$ and all orders in $H_G$
\begin{equation}\begin{split}\label{eq:app1}
I^{(1)}_{\rm gated}&(x,t) = \langle in| T_K \Big[ \hat{I}(x,t^\xi) \, \times \\
-i \sum_\eta \eta & \int dt_1 H_T(t_1^\eta) e^{-i \sum_{\eta'} \eta' \int dt_1' H_G(t_1'^{\eta'})} \Big] |in \rangle.
\end{split}\end{equation}

In terms of bosonic fields:
\begin{equation}\begin{split}\label{eqC2}
I^{(1)}_{\rm gated}(x,t) &= i e \frac{\sqrt{\nu}}{2\pi} \partial_x \sum_\eta \eta \int dt_1 \frac{\Gamma}{2\pi a} \sum_\epsilon e^{-i \epsilon \kappa} \\
&\quad \times \langle 0| T_K \Big[ \phi(x,t^\xi) \, e^A \Big] |0 \rangle
\end{split}\end{equation}
with the shortcut of notation
\begin{equation}\begin{split}\label{exprA}
&A = i\epsilon \sqrt{\nu} \bigl[\phi(d/2,t_1^\eta)-\phi(-d/2,t_1^\eta)\bigr]
\\&
- i\sqrt{\nu} \sum_n \sum_\zeta \zeta \phi(-\mathcal{T},(-\mathcal{T}+\tau_n)^\zeta)
\\&
+ i\frac{e \sqrt{\nu} V_G}{2\pi} \sum_{\eta'} \eta' \int dt_1' (\phi(d/2,t_1'^{\eta'})-\phi(-d/2,t_1'^{\eta'})).
\end{split}\end{equation}
$A$ being linear in $\phi$, we can use Eq.~\eqref{eq:identity} to compute
\begin{equation}\begin{split}\label{eqC4}
    \langle 0| T_K \Big[ \phi(x,t^\xi) \, e^A \Big] |0 \rangle = \langle 0| T_K \Big[ \phi(x,t^\xi) A \Big] |0 \rangle&
    \\\times\exp\left[\frac{1}{2}\langle0|T_K(A^2)|0\rangle\right]&
\end{split}\end{equation}
In the argument of the exponential above, some terms vanish thanks to Eq.~\eqref{eq:UsefulIdentity2}. The remaining one are
\begin{widetext}
\begin{equation}\begin{split}
&\frac{1}{2}\langle0|T_K(A^2)|0\rangle = \nu_\delta\mathcal{G}(d)
+\epsilon\nu_\lambda\sum_nG^r(d/2,t_1;-\mathcal{T},-\mathcal{T}+\tau_n)-G^r(-d/2,t_1;-\mathcal{T},-\mathcal{T}+\tau_n)
\\&
- \epsilon\frac{e\nu_cV_G}{2\pi} \int dt_1' \Big[ G^r(d/2,t_1;d/2,t_1') - G^r(-d/2,t_1;d/2,t_1') - G^r(d/2,t_1;-d/2,t_1') + G^r(-d/2,t_1;-d/2,t_1') \Big]
\end{split}\end{equation}
\end{widetext}
The terms of the first line are familiar since we already encountered them in Eq.~\eqref{eq34}. Although, the second line term is new. Using the expression of $G^r$ given in Eq.~\eqref{eqGr}, it becomes :
\begin{equation}\begin{split}
- \epsilon\frac{e\nu_cV_G}{2\pi} \int dt_1' \Big[ & G^r(d/2,t_1;d/2,t_1') - ... \Big]
\\&=i\epsilon e\nu_cV_G \int dt_1'\;F(t_1-t_1').
\end{split}\end{equation}
With Eq.~\eqref{F_int}, the integral over $t_1'$ reduces to
\begin{equation}
\int dt_1'\;F(t_1-t_1') = \frac{d}{1+c_E},
\end{equation}
If we re-inject in Eq.~\eqref{eqC4} then Eq.~\eqref{eqC2},
\begin{equation}\begin{split}\label{eqC2}
&I^{(1)}_{\rm gated}(x,t) = i e \frac{\sqrt{\nu}}{2\pi} \partial_x \sum_\eta \eta \int dt_1 \frac{\Gamma}{2\pi a} \sum_\epsilon e^{-i \epsilon \kappa} \\
&\quad \times \langle 0| T_K \Big[ \phi(x,t^\xi) A \Big] |0 \rangle e^{\nu_\delta\mathcal{G}(d)}
\\&\quad \times e^{\epsilon\nu_\lambda\sum_nG^r(d/2,t_1;-\mathcal{T},-\mathcal{T}+\tau_n)-G^r(-d/2,t_1;-\mathcal{T},-\mathcal{T}+\tau_n)}
\\&\quad \times e^{i\epsilon \frac{e\nu_cV_Gd}{1+c_E}},
\end{split}\end{equation}
The only terms depending on $\eta$ are now in $\langle 0| T_K \Big[ \phi(x,t^\xi) A \Big] |0 \rangle$.
\begin{equation}\begin{split}
    \sum_\eta\eta\;&\langle 0| T_K \Big[ \phi(x,t^\xi) A \Big] |0 \rangle=\\&i\epsilon\sqrt{\nu}\Bigl[G^r(x,t;d/2,t_1)-G^r(x,t;-d/2,t_1)\Bigr].
\end{split}\end{equation}
Gathering all together, we find the expression
\begin{equation}\begin{split}
&I^{(1)}_{\rm gated}(x,t) = \frac{-e\nu_c\Gamma}{4\pi^2a} e^{\nu_\delta\mathcal{G}(d)}\int dt_1  \sum_\epsilon\epsilon e^{-i \epsilon \kappa+i\epsilon \frac{e\nu_cV_Gd}{1+c_E}}
\\&\quad \times \partial_x\bigl[G^r(x,t;d/2,t_1)-G^r(x,t;-d/2,t_1)\bigr]
\\&\quad \times e^{\epsilon\nu_\lambda\sum_nG^r(d/2,t_1;-\mathcal{T},-\mathcal{T}+\tau_n)-G^r(-d/2,t_1;-\mathcal{T},-\mathcal{T}+\tau_n)}
\end{split}\end{equation}
which is identical to Eq.~\eqref{eq34}, except that a contribution from the gate voltage has been added to the phase
\begin{equation}
\varphi_{\rm ungated} = -\kappa \quad\rightarrow\quad \varphi_{\rm gated} = -\kappa+\frac{e \nu_c V_G d}{1 + c_E}.
\end{equation}

\bibliography{AnyonBiblio}
\end{document}